\newif\ifmnras
\mnrastrue

\ifmnras

\documentclass[a4paper,fleqn,usenatbib]{mnras}
\usepackage{txfonts}
\usepackage[T1]{fontenc}
\usepackage{ae,aecompl}

\usepackage{graphicx}	
\usepackage[section] {placeins}
\usepackage{subfig}
\usepackage{float}
\usepackage{color}
\usepackage{hyperref}
\usepackage{float}

\def\msun{{\rm\,M_\odot}} 
\def\lsun{{\rm\,L_\odot}}
\def\zsun{{\rm\,Z_\odot}}
\newcommand{\etal}{et al.\ }
\newcommand{\kms}{\, {\rm km\, s}^{-1}}
\newcommand{\ikms}{(\kms)^{-1}}
\newcommand{\mpc}{\, {\rm Mpc}}
\newcommand{\kpc}{\, {\rm kpc}}
\newcommand{\hmpc}{\, h^{-1} \mpc}
\newcommand{\ihmpc}{(\hmpc)^{-1}}
\newcommand{\hkpc}{\, h^{-1} \kpc}
\newcommand{\lya}{Ly$\alpha$}
\newcommand{\lyaf}{Ly$\alpha$ forest}
\newcommand{\ch}{\bf change}
\newcommand{\gmo}{{\gamma-1}}
\newcommand{\bF}{\bar{F}}
\newcommand{\hi}{\mbox{H\,{\scriptsize I}\ }}
\newcommand{\heii}{\mbox{He\,{\scriptsize II}\ }}
\newcommand{\civ}{\mbox{C\,{\scriptsize IV}\ }}
\newcommand{\kpa}{k_\parallel}
\newcommand{\vk}{{\mathbf k}}
\newcommand{\df}{\delta_F}
\newcommand{\sF}{{F_s}}
\newcommand{\sdelta}{{\delta_s}}
\newcommand{\seta}{{\eta_s}}
\newcommand{\dt}{\Delta \theta}
\newcommand{\dv}{\Delta v}
\newcommand{\pa}{\parallel}
\newcommand{\pe}{\perp}
\newcommand{\dz}{\Delta z}
\newcommand{\llya}{L$_{{\rm Ly}\alpha}$}
\newcommand{\lheii}{L$_{{\rm He II}}$}
\newcommand{\lciv}{L$_{{\rm C IV}}$}
\newcommand{\expZ}{\langle Z \rangle}
\newcommand{\expT}{\langle T \rangle}
\newcommand{\expD}{\langle n_{{\rm H}} \rangle}
\newcommand{\expF}{\langle f_{{\rm HI}} \rangle}

\title{C {\sc iv} and He {\sc ii} Line Emission of Lyman Alpha Blobs: Powered by Shock Heated Gas}

\author[]{Samuel H. C. Cabot$^{1}$, Renyue Cen$^{1}$ and Zheng Zheng$^{2}$
\\
$^{1}$Department of Astrophysical Sciences, Princeton University, Princeton, NJ 08544\\
$^{2}$University of Utah, Department of Physics and Astronomy, Salt Lake City, UT 84112\\
} 

\pubyear{2016}

\begin{document}
\label{firstpage}
\pagerange{\pageref{firstpage}--\pageref{lastpage}}
\maketitle

\else

\pdfoutput=1
\documentclass[12pt,preprint]{aastex}

\usepackage[T1]{fontenc}
\usepackage{ae,aecompl}

\usepackage{graphicx}	
\usepackage[section] {placeins}
\usepackage{subfigure}
\usepackage{float}
\usepackage{color}
\usepackage{hyperref}
\graphicspath{{./figures/}}
\usepackage{float}

\usepackage[scaled]{helvet}
\renewcommand*\familydefault{\sfdefault}
\usepackage[T1]{fontenc}

\def\msun{{\rm\,M_\odot}} 
\def\lsun{{\rm\,L_\odot}}
\def\zsun{{\rm\,Z_\odot}}
\newcommand{\etal}{et al.\ }
\newcommand{\kms}{\, {\rm km\, s}^{-1}}
\newcommand{\ikms}{(\kms)^{-1}}
\newcommand{\mpc}{\, {\rm Mpc}}
\newcommand{\kpc}{\, {\rm kpc}}
\newcommand{\hmpc}{\, h^{-1} \mpc}
\newcommand{\ihmpc}{(\hmpc)^{-1}}
\newcommand{\hkpc}{\, h^{-1} \kpc}
\newcommand{\lya}{Ly$\alpha$}
\newcommand{\lyaf}{Ly$\alpha$ forest}
\newcommand{\ch}{\bf change}
\newcommand{\gmo}{{\gamma-1}}
\newcommand{\bF}{\bar{F}}
\newcommand{\hi}{\mbox{H\,{\scriptsize I}\ }}
\newcommand{\heii}{\mbox{He\,{\scriptsize II}\ }}
\newcommand{\civ}{\mbox{C\,{\scriptsize IV}\ }}
\newcommand{\kpa}{k_\parallel}
\newcommand{\vk}{{\mathbf k}}
\newcommand{\df}{\delta_F}
\newcommand{\sF}{{F_s}}
\newcommand{\sdelta}{{\delta_s}}
\newcommand{\seta}{{\eta_s}}
\newcommand{\dt}{\Delta \theta}
\newcommand{\dv}{\Delta v}
\newcommand{\pa}{\parallel}
\newcommand{\pe}{\perp}
\newcommand{\dz}{\Delta z}
\newcommand{\llya}{L$_{{\rm Ly}\alpha}$}
\newcommand{\lheii}{L$_{{\rm He II}}$}
\newcommand{\lciv}{L$_{{\rm C IV}}$}
\newcommand{\expZ}{$\langle Z \rangle$}
\newcommand{\expT}{$\langle T \rangle$}
\newcommand{\expD}{$\langle n_{{\rm H}} \rangle$}
\newcommand{\expF}{$\langle f_{{\rm HI}} \rangle$}

\title{C {\sc iv} and He {\sc ii} Line Emission of Lyman Alpha Blobs: Powered by Shock Heated Gas}
\author{Samuel H. C. Cabot$^{1}$, Renyue Cen$^{1}$ and Zheng Zheng$^{2}$} 
\begin{document}
\label{firstpage}

\fi

\begin{abstract}

Utilizing {\it ab initio} ultra-high resolution hydrodynamical simulations, we investigate 
the properties of the interstellar and circumgalactic medium of Ly$\alpha$ Blobs (LABs) at $z=3$,
focusing on three important emission lines: Ly$\alpha$ 1216\AA, \heii 1640\AA\ and \civ 1549\AA. 
Their relative strengths provide a powerful probe of the thermodynamic properties of the 
gas when confronted with observations. 
By adjusting the dust attenuation effect using one parameter 
and matching the observed size-luminosity relation of LABs using another parameter,
we show that our simulations can reproduce the observed \civ/\lya\ and \heii/\lya\ ratios adequately.
This analysis provides the first successful physical model to account for simultaneously 
the LAB luminosity function, luminosity-size relation, and the \civ/Ly$\alpha$ and \heii/Ly$\alpha$ ratios,
with only two parameters.
The physical underpinning for this model is that,
in addition to the stellar component for the \lya\ emission,
the \lya\ and \civ\ emission lines due to shock heated gas are primarily
collisional excitation driven and the \heii\ emission line collisional ionization driven. 
We find that the density, temperature and metallicity of the gas responsible for each emission line
is significantly distinct, in a multi-phase interstellar and circumgalactic medium that
is shock-heated primarily by supernovae and secondarily by gravitational accretion of gas.
\end{abstract}

\begin{keywords}
methods: numerical,
galaxies: clusters: general,
galaxies: starburst,
quasars: emission lines
\end{keywords}


\section{Introduction}

\lya\ blobs (LABs) \citep[e.g.,][]{1999Keel,2000Steidel,2004Matsuda} are 
the largest and most luminous \lya\ emitters in the universe, 
with luminosities of \llya $ > 10^{42.5} {\rm ergs \: s}^{-1}$ and 
extending up to hundreds of kiloparsecs in size. 
In the context of galaxy formation, these $z \sim 2-6$ objects are important
because they lie in overdense regions, and tend to be proto-galaxy 
clusters \citep[e.g.,][]{2010Yang,2013dCen}.
However, the mechanisms which power the \lya\ emission of these sources are debated. 
\citet{2013dCen} suggest that central stellar emission (and possibly central AGN) 
is the primary source of power, 
with gravitational cooling radiation being the secondary source, in their star-burst model (SBM). 
The large spatial extent of LABs is a combination of the contribution from clustered sources (galaxies)
and spatial diffusion of \lya\ photons through resonant scatterings by circumgalactic medium.
With detailed radiative transfer calculations, 
this model is the only model that is able to reproduce both the observed 
luminosity-size relation and luminosity function of LABs \citep{2004Matsuda,2011Matsuda}.

Recently, \citet{2014Cantalupo} discover one of the largest LABs 
that contains a luminous quasar, UM 287.
In a novel model they propose that the observed \lya\ emission could be explained by 
a high clumping factor ($C \approx 1000$), corresponding to a set of small, cold 
gas clumps illuminated and photoionized by the quasar.
\citet{2015aBattaia} turn to a sample of 13 LABs and explore the possibility that such 
photoionization is the primary power source of LABs via deep observations of \heii and \civ emission lines. 
Their non-detections of these lines and upper limits on 
line ratios are reproduced in photoionization models: the optically thin scenario 
requires $n_{\rm H} \gtrsim 6 {\rm cm}^{-3}$; the optically thick scenario requires 
a weak AGN and $N_{\rm H} > 17.2 {\rm cm}^{-2}$. These results agree with 
\citet{2014Cantalupo}. 
\citet{2015bBattaia} return to the case of UM 287, seeking to break the degeneracy between 
clumpiness and the total amount of cool gas (a low clumping factor requires unreasonably 
high mass for the dark matter halo). Despite extremely deep observations, they again 
fail to detect the \heii and \civ emission lines. They rule out the optically thick scenario, since 
it would require too high a \lya\ surface brightness (or a covering factor so small that it conflicts 
with the observed morphology), and thus turn to optically thin models. 
They find that the \lya\ emission is due to dense clumps
($n_{\rm H} \gtrsim 3 {\rm cm}^{-3}$ )
with a radius of $R \lesssim 20$ pc, also in agreement with 
\citet{2014Cantalupo}.

These physical constraints are so strict, however, that it prompts 
search of alternative models. Indeed, \citet{2015aBattaia} discuss several other candidates for 
the primary power mechanism, including shocks, gravitational cooling radiation, and resonant 
scattering (AGN may remain an additional power source, but a non-essential one).
Here we use ultra-high resolution simulations to explore the successful starburst model of LABs
put forth in \citet{2013dCen},
where the \lya\ emission is primarily powered by central starbursts (and possibly AGN),
not due to fluorescence of dense clouds illuminated by quasars.
We compute \heii and \civ emission of the same LABs analyzed in \citet{2013dCen}
in addition to the \lya\  line.
We perform a detailed comparison between the model predictions and 
extant observational data in \citet{2015aBattaia}, \citet{2015bBattaia} 
and other previous observational studies of LABs \citep[e.g.,][]{2005Dey,2009Prescott}. 
We find that the \heii and \civ emission lines in our model
- powered by a combination of both collisional ionization due to shocks and photoionization by resident stars -
provide an adequate match to observations with respect to the 
joint \heii/\lya\ and \civ/\lya\ emission line ratios in the LABs.
The success is largely due to the multiphase nature of the CGM in galaxies and 
a substantial amount of shock heated gas that has favorable temperatures and densities to make
large contributions to the line emission in question.
It is also relevant to note that an adequate implementation of supernova feedback process
is crucial to redistributing metals in the multiphase medium so as to produce the right amount of \civ emission
and other metals lines, such as the adequate reproduction 
of the Si II absorption line width of damped Lyman alpha systems
\citet{2012Cen} and O VI absorption lines in the intergalactic medium \citet{2012bCen},
two exemplary sets of observables spanning a significant range in redshift, gas density, temperature and environment.
This paper is structured as follows: In \S 2 we give description of the simulation and 
our analysis method.
A detailed comparison between simulation results and observational data is in \S 3.
We discuss the physical picture of the emission mechanism in \S 4. Our conclusions are summarized in \S 5.

\section{Simulations and Methods}

\subsection{Cosmological Hydrodynamic Simulations}

For a more detailed description of the {\it ab initio} large-scale adaptive mesh-refinement ominiscient zoom-in ({\color{red}\bf LAOZI}) 
simulations, performed with the adaptive mesh-refinement (AMR) code, Enzo \citep[][]{2014Bryan},
see \citet[][]{2014Cen}. 
Briefly, we use the WMAP7-normalized \citep[][]{2011Komatsu} $\Lambda$CDM model:
$\Omega_M=0.28$, $\Omega_b=0.046$, $\Omega_{\Lambda}=0.72$, $\sigma_8=0.82$,
$H_0=100 h \,{\rm km\, s}^{-1} {\rm Mpc}^{-1} = 70 \,{\rm km\, s}^{-1} {\rm Mpc}^{-1}$ and $n=0.96$.
A zoom-in box of size $21\times 24\times 20h^{-3}$Mpc$^3$ comoving is embedded in a $120~h^{-1}$Mpc periodic box.
The maximum resolution in the zoom-in box is better than $111h^{-1}$pc (physical) at all times.
Star formation follows the prescription of \citet[][]{1992CenOstriker}.
Supernova feedback from star formation is modeled following \citet[][]{2005Cen}. 
The \citet{2012Haardt} UV radiation background is used in the simulation 
along with a treatment of self-shielding \citet{2005Cen}.
This is the same simulation used in the successful demonstration of a starburst based model for the LABs \citep{2013dCen},
coupled with detailed radiative transfer calculations of \lya\ radiation 
and the successful interpretation of the observed diffuse Lyman-alpha halos in Lyman-alpha emitting
galaxies \citep{2015Lake}.
Since the goal of the simulations is to model shock flows, 
AGN feedback is not included (though could be added).

Shock velocities are due to gravitational interactions and stellar feedback,
and velocities range from about 100km s$^{-1}$ to 1000km s$^{-1}$.
If the shock velocities are too high or too low for a given species,
that particular shock would not contribute much;
the abundant species are shocked to the correct temperatures
by shocks of appropriate velocities. 
These velocities cover both the low velocity outflows 
observed by \citep{2014bYang}, and accommodate large-scale outflows caused by strong shocks as predicted by
\citet{2000Taniguchi} and \citet{2006Mori}; they are also consistent with the velocities
from the shock model in \citet{2015aBattaia} that explain the observed line ratios ($> 250$km s$^{-1}$). 
While 1000km s$^{-1}$ outflows are not excluded from our 
model, we expect line emission to be primarily dominated by internal shock heated 
gas, driven by both supernova feedback and gravitational accretion. These processes may explain the observational bias of 
lower velocities \citep{2011Yang, 2014bYang} and are consistent with the findings 
of \citet{2013McLinden}.

\subsection{Analysis Method}

\ifmnras

\begin{figure}
\centering
	\includegraphics[width=78mm]{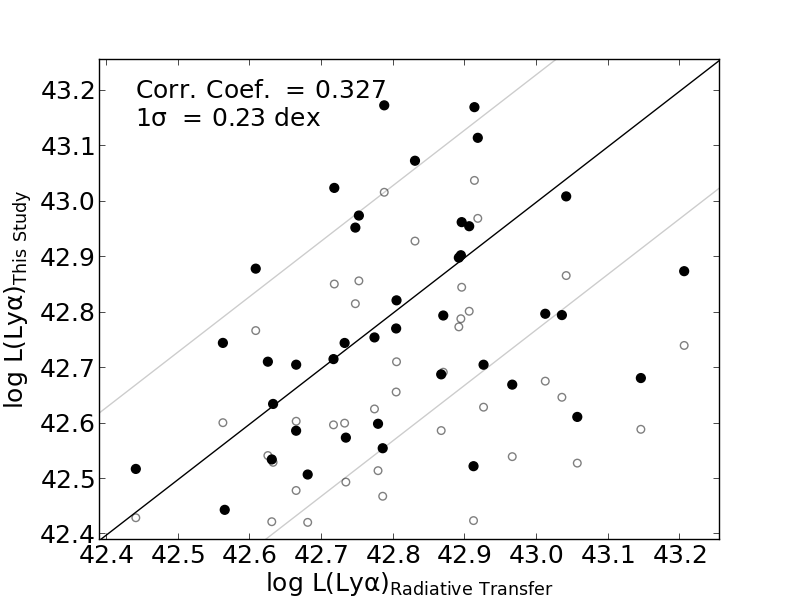}
    \caption{
Our \lya\ luminosity plotted against that obtained from detailed radiative transfer.
1$\sigma$ deviation lines are plotted surrounding the equality line.
The parameter value $\gamma = 6$ is used to generate our values shown here (filled points).
For comparison, we also show the poor scatter and 
downward shift generated by $\gamma = 7.5$ (open points).
}
    \label{fig:lvl}
\end{figure}

\else

\begin{figure}[!ht]
\centering
	\includegraphics[width=150mm]{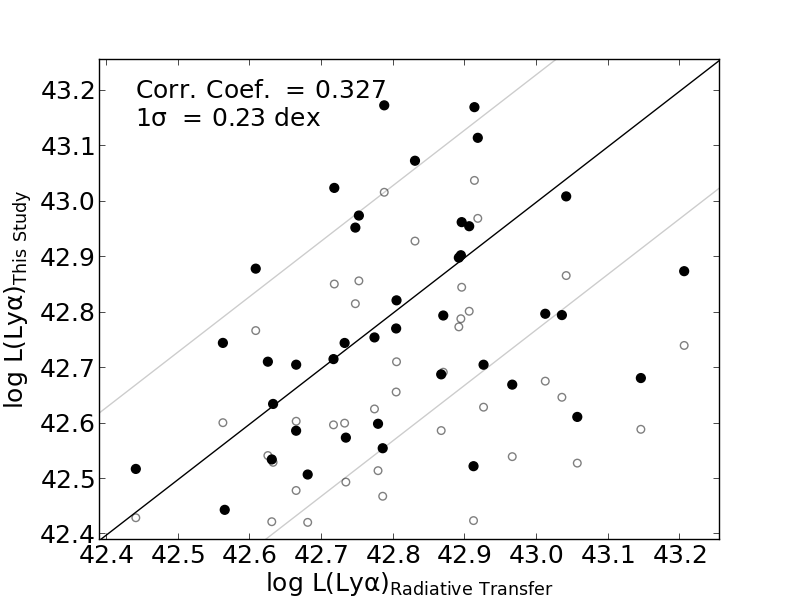}
    \caption{
Our \lya\ luminosity plotted against that obtained from detailed radiative transfer.
1$\sigma$ deviation lines are plotted surrounding the equality line.
The parameter value $\gamma = 6$ is used to generate our values shown here (filled points).
For comparison, we also show the poor scatter and 
downward shift generated by $\gamma = 7.5$ (open points).
}
    \label{fig:lvl}
\end{figure}

\fi

\ifmnras

\addtolength{\tabcolsep}{-5pt}
\begin{figure}
\centering
		\includegraphics[width=83mm]{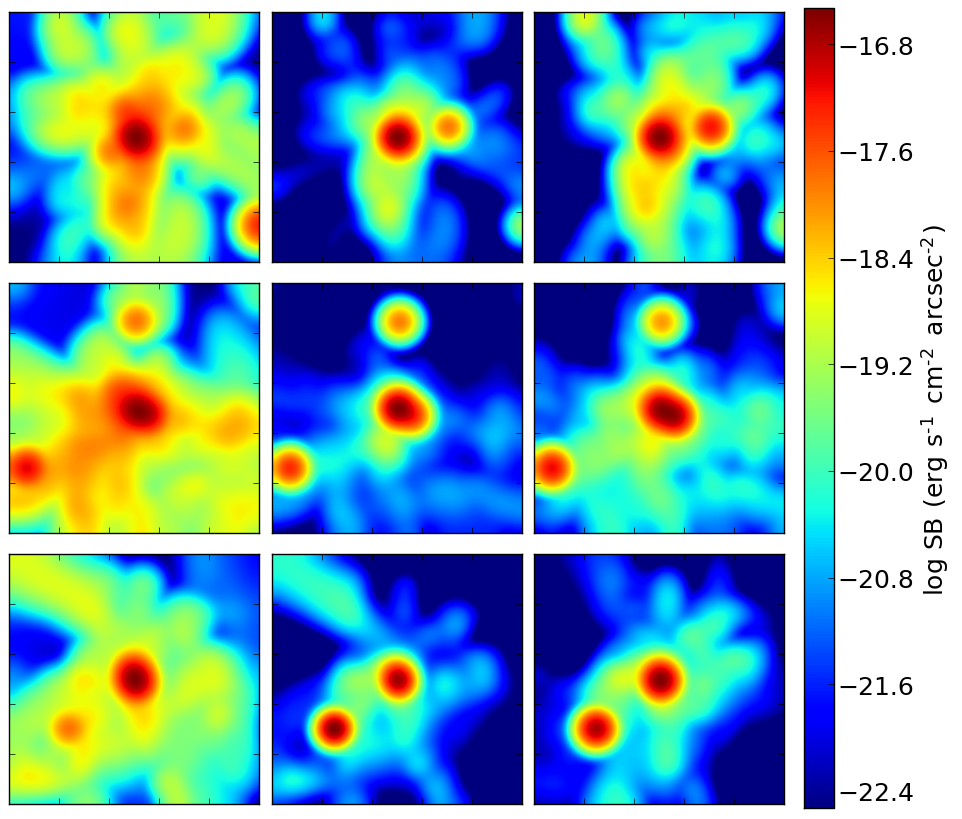}
    \caption{
Surface brightness maps of size $100\times 100$kpc$^2$ 
for three randomly selected galaxies with ID: 0 (top row), 20 (center row), and 42 (bottom row), 
projected along a depth of $100$kpc.
The maps are in each emission wavelength: \lya\ (left column), \heii (center column) and \civ (right column). 
A gaussian blur of full width at half maximum (FWHM) = 1" was applied to the images to simulate seeing 
conditions in \citet{2014Cantalupo} (FWHM = 0.6 - 1.0") and \citet{2015aBattaia} (0.5 - 1.4").
}
    \label{tab:maps}
\end{figure}
\addtolength{\tabcolsep}{5pt}

\else

\addtolength{\tabcolsep}{-5pt}
\begin{figure}[!ht]
\centering
		\includegraphics[width=0.9\textwidth]{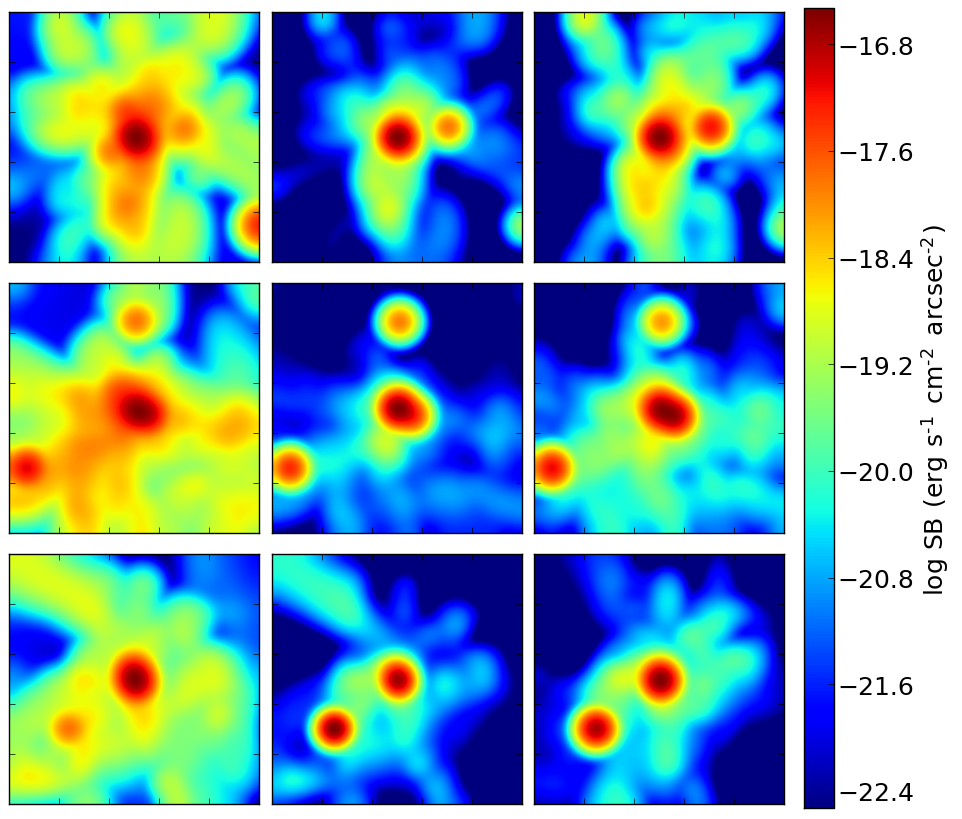}
    \caption{
Surface brightness maps of size 100 kpc x 100 kpc 
for three randomly selected galaxies with ID: 0 (top row), 20 (center row), and 42 (bottom row), 
projected along a depth of $100$kpc.
The maps are in each emission wavelength: \lya\ (left column), \heii (center column) and \civ (right column). 
A gaussian blur of full width at half maximum (FWHM) = 1" was applied to the images to simulate seeing 
conditions in \citet{2014Cantalupo} (FWHM = 0.6 - 1.0") and \citet{2015aBattaia} (0.5 - 1.4").
}
    \label{tab:maps}
\end{figure}
\addtolength{\tabcolsep}{5pt}

\fi

Our analysis method consists of three steps.
First, we utilize the spectral synthesis code Cloudy version 13.03 \citep{2013Ferland} 
to generate emission spectra (the transmitted continuum component, in Cloudy) 
over a range of gas density and temperature at solar metallicity. 
Line emissions for \lya, \civ and \heii are tabulated 
for a two-dimensional grid spanning density ${\rm n_{H}} = 10^{-4} - 10^{4} {\rm cm}^{-3}$ and temperature
T$ = 10^{3} - 10^{6} {\rm K}$, both with steps of 0.025 dex. 
We use the \citet{2012Haardt} UV background incident radiation field at $z=3.1$ for the calculation, 
subject to self-shielding effect of each individual region based on local optical depth calculations.

Second, we select 40 most massive galaxies in our simulation at $z= 3.1$, 
the same sample used in the LAB model paper \citep{2013dCen}.
We then use the analysis software yt version 2.6 \citep{2011Turk}
to perform the following for each galaxy: 

\begin{enumerate}
\item We identify a cubic region of length 100 kpc centered on the galaxy, 
which covers approximately the area of the $2.2 \times 10^{-18} {\rm \: ergs \: s^{-1} \: cm^{-2} \: arcsec^{-2}}$ \lya\ isophotal region in \citet{2015aBattaia}, 
and \citet{2004Matsuda}.

\item We obtain physical variables, including density, temperature, metallicity, and self-shielding optical depth 
on a unigrid covering the 100kpc cube at the resolution of 160 pc. 
The HM background is subject to self-shielding attenuation, 
and is computed as $\varepsilon=\varepsilon_{0} \times \langle \exp(-n_{HI} h \sigma_\lambda) \rangle_{{\rm cell}}$,
where $\varepsilon_{0}$ is unattenuated emission,
$n_{HI}$ is the neutral hydrogen density of the cell,
$h$ is the local scale-height along each of the six faces of the cell
and $\sigma_\lambda$ is the mean absorption cross section of ionizing photons given the spectral shape of the HM background.
The attenuation average "$<>_{{\rm cell}}$" is done over the six faces of the cell.

\item We use the Cloudy generated emissivity lookup tables to obtain line emissions for each grid cell (of size 160pc) within the 100kpc box.
The metallicity parameter is a multiplicative factor since our table depends 
only on temperature and density with the assumed solar metallicity.

\item We add the stellar contribution to the \lya\ emssion in each grid cell,
and use a scaling relation that a star formation rate (SFR) 
of $1 \msun / {\rm yr}$ produces \llya = $1 \times 10^{42} {\rm ergs\:s^{-1}}$
\citep{2010Zheng, 2013dCen}.

\item We model dust obscuration for the emission 
as follows (this is for dust on small scales within the galaxies; 
while there may be dust on $\sim 100$kpc scales \citep[e.g.,][]{2014Peeples}, we do not assume so).
The optical depth at wavelength $\lambda$ along each of the six faces of each grid cell is
\begin{equation}
    \tau_{\lambda} = 0.49 \times \frac{Z}{\zsun} \frac{N_{{\rm H}}}{10^{21}} \frac{A_{\lambda}}{A_{\nu}}\beta (1 + \gamma)
	\label{eq:absorption}
\end{equation}
\end{enumerate}
where $A_{\nu} =  4.89 \times 10^{-22}$ and $A_{\lambda}$ is the absorption coefficient at the emission line wavelength 
\citep{2003Draine}, 
$Z$ is gas metallicity of the cell,
$\beta$ is to account for the uncertainties in metal (hence dust) density modeling uncertainty in the simulation and 
$\gamma$ is zero for non-resonant lines (\heii)
and nonzero to account for scatterings of resonant lines (\lya\ and \civ).
In our case, to be self-consistent, a single $\gamma$ applies to both \lya\ and \civ lines.
Note that we only have two free parameters in our calculations, with four outputs -
\lya\ luminosity, \lya\ luminosity-size relation, \civ/\lya\ ratio and \heii/\lya\ ratio.
The metallicity-weighted column density $N_{\rm H}$ is determined by computing the scale-height 
of the metallicity-weighted neutral hydrogen (volumetric) density along each of the six faces of each grid cell,
then multiplied by the hydrogen volumetric density of the cell. This is done using a smoothed covering grid, 
which interpolates coarse regions of the simulation to match high resolution dimensions.
The mean transmitted emission $<{\rm L}>$ for each emission line from each grid cell is computed using
\begin{equation}
    <{\rm L}> = {\rm L} <e^{-\tau_{\lambda}}>,
	\label{eq:attenuation}
\end{equation}
averaged over six faces of each grid cell, where L is the intrinsic line emission, including contribution from 
both shock heated gas and stars.

Let us now discuss the purpose and logic of having two parameters,
$\beta$ and $\gamma$ (Eq~\ref{eq:absorption}).
We first note that we have three separate emission lines and only two free parameters in our calculation.
Since we do not perform detailed, expensive radiative transfer calculations
for the \civ line that is a resonant line,
we use the parameter $\gamma$ in Eq~\ref{eq:absorption} as a proxy method, in the following sense.
The parameter $\gamma$ (or more precisely, $\beta (1+\gamma)$)
is adjusted to reproduce, within the scatter,
the \lya\ emission obtained through detailed radiative transfer calculations performed in \citet{2013dCen}.
We use the same $\gamma$ for the \civ line.
That leaves us with only one free parameter $\beta$ that we vary, admittedly by hand, 
with the sole purpose of searching for a match to the observations
in the 
\heii/\lya\ -\civ/\lya\  
line ratios 
plane.
Since the observable numbers outnumber the adjustable parameters here,
a successful outcome from this ``fitting" exercise represents a non-trivial result.
It is noted, though, that if the \civ emission is lower than required to yield reasonable fit,
there would be no adequate match to observations.

\ifmnras

\begin{figure}
\centering
	\includegraphics[trim={1.2cm 0.7cm 1.8cm 2.0cm},clip,width=86mm]{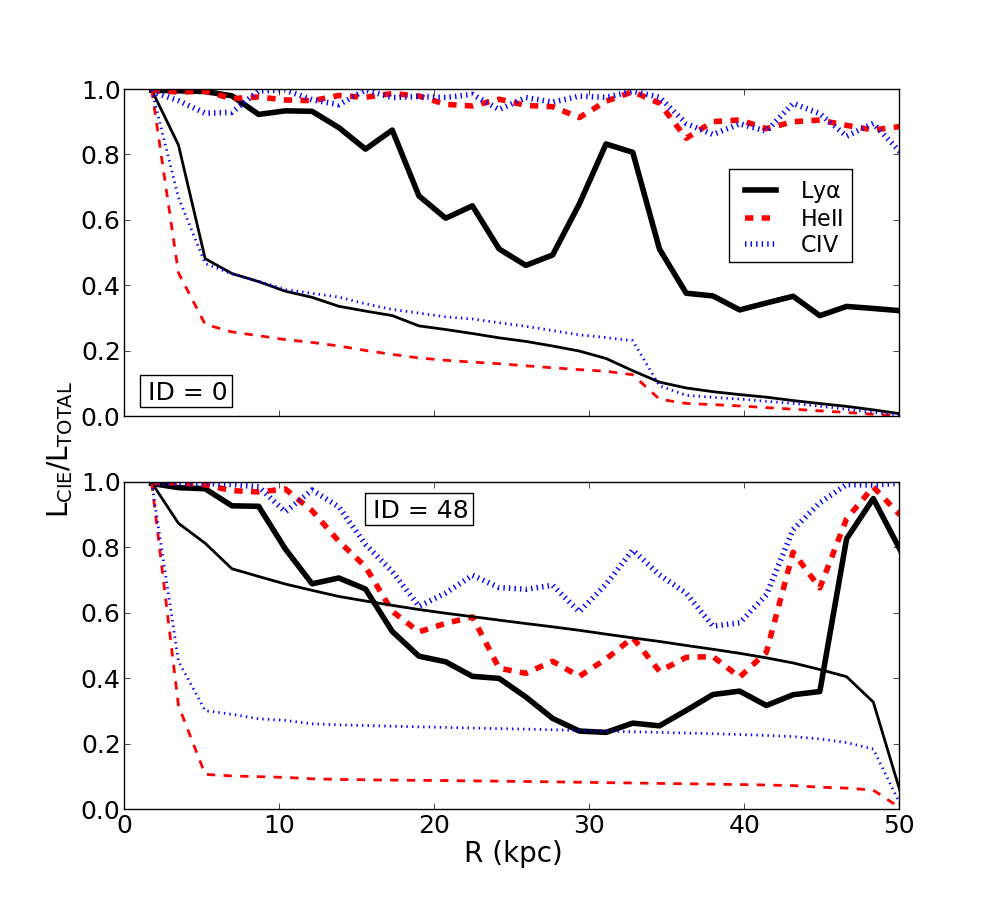}
    \caption{
Radial profiles of the fractional luminosity produced 
by collisional processes for \lya\ (solid black line), \heii (dashed red line), 
and \civ (dotted blue line).
Thick lines represent the differential profile, and thin lines are the 
normalized cumulative total luminosity. The latter demonstrate 
that most of the emission is concentrated in the central regions.
The panels represent galaxies of two very different stellar mass regimes: 
$4.33 \times 10^{11}$M$_{\odot}$ (Top) and 
$5.43 \times 10^{10}$M$_{\odot}$ (Bottom).}
    \label{fig:prof}
\end{figure}

\else

\begin{figure}
\centering
	\includegraphics[trim={1.2cm 0.7cm 1.8cm 2.0cm},clip,width=150mm]{RADPROF.png}
    \caption{
Radial profile of the fractional luminosity produced 
by collisional processes. Emission lines are depicted as black circles, 
red squares, and yellow triangles for \lya, \heii, and \civ respectively.
Solid markers depict the differential profile and hollow markers are cumulative.
The panels represent galaxies of two very different stellar mass regimes: 
$4.33 \times 10^{11}$M$_{\odot}$ (Left) and 
$5.43 \times 10^{10}$M$_{\odot}$ (Right).}
    \label{fig:prof}
\end{figure}

\fi

We find that parameters within $\pm \sim25\%$ of $\beta = 2/7$ and $\gamma=6$ are able to provide 
a reasonable match to observations with respect to the \civ/\lya \ ratio and \heii/\lya \ ratio,
as well as being able to reproduce \lya\ luminosities of galaxies 
obtained with detailed \lya\ radiative calculations \citep{2013dCen}. 
The correlation between our \lya\ emission and that from radiative transfer 
is shown in Figure~\ref{fig:lvl}. Also shown is the case of  
$\gamma = 7.5$. We determine our eventual value of $\gamma$ such that the scatter is minimized.
We find that the $1\sigma$ r.m.s variation is $\sim 0.2$ dex 
in $L_{\tau}/L_{RT}$, 
where $L_{\tau}$ and $L_{RT}$ are the \lya\ luminosity computed here and that computed using 
radiative calculations \citep{2013dCen}, respectively; 
the r.m.s. variation is consistent with and within the variation among different viewing angles for a typical galaxy
with detailed radiative transfer calculations \citep{2014Zheng}.
In any case, this small variation does not alter our conclusions reached below.
However, small variations of $10-30\%$ on either of the two parameters ($\beta$ and $\gamma$)
change the results significantly, and in some cases may render the results inconsistent with observations. 

In Figure~\ref{tab:maps} we show surface brightness maps for three randomly selected galaxies.
Overall, the emissions for all three lines follow the galactic structure centered on the galaxy.
In finer details, they display significant differences, which are ultimately due to the fact that
the regions that are responsible for three emission lines are different,
arising from regions of different physical characteristics in a multi-phase medium,
as will be made clear later.
Table~\ref{tab:ratios} summarizes the luminosities of the three lines and two ratios for each of the 40 galaxies in the simulation. 
Contribution from collisional processes is easily separated from the 
UV photoionization/photoexcitation 
component by setting the ionization parameter to 0 for all of the simulation
space. We find that emission from collisional processes traces the total emission (shown for a few galaxies in 
Figure~\ref{tab:maps}), and accounts for the 
vast majority ($> 70\%$) of the total luminosity for each line 
in Table~\ref{tab:ratios}, indicating that the photoionization/photoexcitation contribution is small.
The fractional luminosity produced by collisional processes is
depicted in Figure~\ref{fig:prof}. It is clear that the collisional processes are largely responsible 
for the emission in the central region that dominates the overall emission. In the outer regions, 
however, the photoionization/excitation processes make a significant, sometimes larger, contribution.
Note that the ratios reported 
here are derived from emission 
over the entire 100kpc box, with no surface brightness threshold. Because our emission 
model does not use radiative transfer, we are unable to produce isophotes 
or sizes of the LABs. However, in the next section we discuss an approach 
to approximate effects of SB thresholds using previous radiative transfer data,
which allows fair comparisons to observations.

\section{Comparison between Simulation Results and Observations}

\setlength\tabcolsep{3.9pt}

\begin{table*}
	\small
 	\centering
 	\caption{Star formation rate, stellar mass, halo mass and luminosities over the 100kpc cube for each galaxy in our sample, and the corresponding ratios. 
	Also emission-weighted metallicity, density and temperature. 
	Column 2 [$\msun\:{\rm yr}^{-1}$], 
	Column 3-4 [$10^{10}\:\msun$], 
	Column 5-8 [$10^{42}{\:\rm ergs\:s^{-1}}$]
	Column 11-13 [$\zsun$], 
	Column 14-16 [${\rm cm}^{-3}$], 
	Column 17-19 [$10^{4}\:{\rm K}$].}
 	\label{tab:ratios}
 	\begin{tabular}{lrrrrcccccccccrrcrr} 
 		\hline
 		ID & SFR & M* & M$_{{\rm Halo}}$ & \llya & \llya* & \lheii & \lciv & $\frac{{\rm He II}}{{\rm Ly}\alpha}$ & $\frac{{\rm C IV}}{{\rm Ly}\alpha}$ & ${\expZ \atop {{\rm (Ly}\alpha \rm )}}$ & ${\expZ \atop {{\rm (He II)}}}$ & ${\expZ \atop {{\rm (C IV)}}}$ & ${\expD \atop {{\rm (Ly}\alpha \rm )}}$ & ${\expD \atop {{\rm (He II)}}}$ & ${\expD \atop {{\rm (C IV)}}}$ & ${\expT \atop {{\rm (Ly}\alpha \rm )}}$ & ${\expT \atop {{\rm (He II)}}}$ & {$\expT \atop {{\rm (C IV)}}$} \\
 		\hline
0 & 409.20 & 43.31 & 397.7 & 11.825 & 5.457 & 0.094 & 0.407 & 0.008 & 0.034 & 0.55 & 0.80 & 0.79 & 3.04 & 14.58 & 2.90 & 2.45 & 24.93 & 11.30 \\
1 & 129.60 & 36.78 & 234.7 & 6.218 & 1.293 & 0.114 & 0.745 & 0.018 & 0.120 & 0.40 & 1.10 & 1.15 & 3.12 & 7.21 & 3.66 & 2.77 & 18.92 & 11.05 \\
2 & 105.40 & 35.57 & 231.4 & 7.565 & 4.444 & 0.075 & 0.970 & 0.010 & 0.128 & 0.71 & 1.46 & 1.58 & 1.20 & 6.47 & 2.52 & 2.39 & 11.59 & 10.65 \\
3 & 113.80 & 29.26 & 192.2 & 10.576 & 4.297 & 0.175 & 0.744 & 0.017 & 0.070 & 0.48 & 0.78 & 0.66 & 6.30 & 15.87 & 5.53 & 2.51 & 13.13 & 10.44 \\
4 & 86.58 & 27.08 & 334.6 & 13.028 & 2.019 & 0.167 & 1.775 & 0.013 & 0.136 & 0.46 & 0.92 & 0.93 & 3.02 & 7.36 & 4.05 & 2.07 & 10.66 & 10.40 \\
5 & 75.82 & 22.57 & 194.4 & 7.922 & 4.083 & 0.038 & 0.371 & 0.005 & 0.047 & 0.61 & 1.04 & 1.16 & 1.59 & 4.90 & 2.56 & 2.28 & 12.68 & 10.57 \\
6 & 87.79 & 20.33 & 175.2 & 14.800 & 5.441 & 0.115 & 1.155 & 0.008 & 0.078 & 0.54 & 1.00 & 1.11 & 2.29 & 5.53 & 2.69 & 2.03 & 9.71 & 10.58 \\
7 & 86.42 & 19.25 & 148.0 & 4.879 & 2.226 & 0.041 & 0.399 & 0.008 & 0.082 & 0.55 & 1.25 & 1.09 & 1.74 & 9.93 & 2.82 & 2.37 & 15.64 & 11.09 \\
8 & 100.20 & 18.31 & 141.0 & 5.080 & 2.204 & 0.056 & 0.204 & 0.011 & 0.040 & 0.46 & 1.30 & 1.09 & 0.70 & 13.53 & 2.51 & 2.22 & 12.82 & 11.24 \\
9 & 38.50 & 17.13 & 137.7 & 6.638 & 2.071 & 0.071 & 0.541 & 0.011 & 0.081 & 0.47 & 1.05 & 1.07 & 2.20 & 6.99 & 3.69 & 2.30 & 9.44 & 10.43 \\
11 & 62.07 & 16.69 & 100.1 & 5.906 & 2.117 & 0.055 & 0.802 & 0.009 & 0.136 & 0.72 & 1.17 & 1.22 & 1.98 & 5.63 & 2.92 & 2.36 & 9.41 & 10.75 \\
12 & 106.30 & 16.59 & 124.8 & 14.914 & 6.686 & 0.137 & 1.011 & 0.009 & 0.068 & 0.62 & 1.00 & 0.99 & 3.75 & 9.08 & 4.33 & 2.08 & 9.13 & 10.52 \\
14 & 71.34 & 15.13 & 139.6 & 9.184 & 3.426 & 0.129 & 0.803 & 0.014 & 0.087 & 0.48 & 1.29 & 1.15 & 2.41 & 12.44 & 4.31 & 2.10 & 9.70 & 10.51 \\
15 & 57.44 & 14.06 & 117.9 & 9.439 & 3.198 & 0.041 & 0.240 & 0.004 & 0.025 & 0.44 & 0.97 & 0.83 & 1.60 & 9.43 & 2.73 & 1.92 & 10.40 & 10.51 \\
16 & 74.77 & 13.92 & 137.9 & 5.142 & 2.100 & 0.119 & 1.261 & 0.023 & 0.245 & 0.76 & 1.11 & 1.11 & 3.86 & 7.75 & 4.00 & 2.46 & 9.68 & 10.75 \\
19 & 37.27 & 12.73 & 115.9 & 8.978 & 1.933 & 0.042 & 0.269 & 0.005 & 0.030 & 0.50 & 1.08 & 0.76 & 1.73 & 12.43 & 2.15 & 2.00 & 10.45 & 10.47 \\
20 & 84.87 & 12.67 & 167.3 & 6.276 & 1.779 & 0.130 & 0.433 & 0.021 & 0.069 & 0.49 & 1.08 & 0.99 & 1.36 & 16.15 & 3.99 & 2.16 & 11.24 & 10.88 \\
21 & 44.42 & 12.16 & 100.5 & 5.562 & 1.441 & 0.047 & 0.321 & 0.009 & 0.058 & 0.40 & 1.01 & 0.98 & 3.24 & 8.55 & 3.40 & 2.11 & 9.51 & 10.54 \\
23 & 79.97 & 10.54 & 98.6 & 4.808 & 0.722 & 0.044 & 0.233 & 0.009 & 0.048 & 0.42 & 1.64 & 1.58 & 0.52 & 10.36 & 3.38 & 1.94 & 9.65 & 10.61 \\
25 & 22.56 & 9.17 & 196.3 & 8.004 & 1.023 & 0.081 & 0.544 & 0.010 & 0.068 & 0.42 & 1.38 & 1.13 & 1.53 & 10.18 & 2.96 & 1.92 & 10.38 & 10.86 \\
26 & 67.92 & 7.76 & 87.1 & 3.593 & 1.017 & 0.040 & 0.178 & 0.011 & 0.050 & 0.41 & 1.69 & 1.41 & 0.57 & 14.46 & 3.14 & 2.00 & 12.89 & 10.65 \\
28 & 43.67 & 7.54 & 80.4 & 5.201 & 1.298 & 0.059 & 0.653 & 0.011 & 0.125 & 0.49 & 1.04 & 1.03 & 2.01 & 6.10 & 4.05 & 2.36 & 9.11 & 10.29 \\
29 & 12.67 & 7.62 & 53.9 & 5.563 & 0.635 & 0.251 & 0.304 & 0.045 & 0.055 & 0.28 & 0.37 & 0.53 & 5.01 & 48.90 & 17.82 & 1.86 & 8.94 & 10.84 \\
30 & 31.54 & 7.42 & 94.8 & 10.221 & 1.295 & 0.029 & 0.342 & 0.003 & 0.033 & 0.52 & 0.78 & 0.96 & 1.92 & 3.61 & 2.02 & 1.96 & 8.20 & 10.49 \\
31 & 14.15 & 7.57 & 61.3 & 3.297 & 0.619 & 0.006 & 0.036 & 0.002 & 0.011 & 0.38 & 0.40 & 0.52 & 0.98 & 1.24 & 0.45 & 2.23 & 13.74 & 10.79 \\
32 & 26.54 & 7.32 & 73.3 & 9.030 & 2.660 & 0.064 & 0.614 & 0.007 & 0.068 & 0.44 & 0.93 & 0.83 & 4.37 & 6.15 & 3.00 & 2.15 & 8.30 & 10.22 \\
33 & 16.78 & 7.02 & 44.0 & 3.222 & 0.785 & 0.018 & 0.114 & 0.005 & 0.035 & 0.35 & 1.38 & 1.39 & 0.78 & 8.31 & 2.41 & 2.20 & 9.95 & 10.66 \\
34 & 25.14 & 6.94 & 102.6 & 6.234 & 0.949 & 0.016 & 0.203 & 0.003 & 0.033 & 0.39 & 0.95 & 1.17 & 0.89 & 3.10 & 2.14 & 1.88 & 9.66 & 10.68 \\
35 & 8.44 & 6.74 & 63.5 & 3.755 & 0.513 & 0.005 & 0.043 & 0.001 & 0.011 & 0.42 & 0.50 & 0.80 & 0.51 & 0.61 & 0.41 & 1.88 & 8.25 & 9.94 \\
37 & 18.92 & 6.65 & 149.3 & 5.082 & 0.638 & 0.059 & 0.250 & 0.012 & 0.049 & 0.55 & 1.44 & 1.14 & 0.92 & 12.80 & 3.04 & 1.96 & 10.64 & 11.02 \\
38 & 18.09 & 6.57 & 104.0 & 4.092 & 0.996 & 0.039 & 0.179 & 0.009 & 0.044 & 0.48 & 1.70 & 1.30 & 0.70 & 14.47 & 3.37 & 2.04 & 11.29 & 10.78 \\
40 & 24.14 & 6.50 & 109.0 & 7.492 & 1.231 & 0.057 & 0.381 & 0.008 & 0.051 & 0.50 & 1.12 & 1.10 & 1.77 & 9.15 & 2.82 & 1.96 & 10.03 & 10.78 \\
41 & 7.26 & 6.38 & 49.2 & 2.782 & 0.566 & 0.006 & 0.068 & 0.002 & 0.024 & 0.35 & 1.43 & 1.75 & 0.97 & 3.12 & 2.04 & 1.93 & 8.58 & 10.37 \\
42 & 17.91 & 6.09 & 57.7 & 3.864 & 0.718 & 0.013 & 0.108 & 0.003 & 0.028 & 0.41 & 1.01 & 1.12 & 1.17 & 3.83 & 3.00 & 2.13 & 9.16 & 10.56 \\
43 & 23.99 & 5.78 & 43.7 & 3.429 & 0.462 & 0.049 & 0.190 & 0.014 & 0.055 & 0.37 & 1.32 & 1.08 & 0.97 & 16.71 & 3.69 & 1.91 & 10.82 & 11.50 \\
44 & 66.83 & 5.81 & 62.7 & 5.691 & 2.199 & 0.115 & 1.082 & 0.020 & 0.190 & 0.56 & 0.98 & 1.04 & 3.02 & 7.74 & 4.46 & 2.38 & 9.00 & 10.73 \\
45 & 37.83 & 5.90 & 52.4 & 4.678 & 1.390 & 0.053 & 0.263 & 0.011 & 0.056 & 0.46 & 1.04 & 1.01 & 2.11 & 10.70 & 4.14 & 2.26 & 9.62 & 10.82 \\
46 & 26.04 & 5.55 & 47.4 & 3.977 & 2.036 & 0.020 & 0.080 & 0.005 & 0.020 & 0.44 & 1.70 & 1.52 & 0.66 & 9.18 & 3.35 & 2.10 & 10.70 & 10.70 \\
47 & 24.20 & 5.39 & 57.2 & 3.335 & 0.370 & 0.040 & 0.291 & 0.012 & 0.087 & 0.36 & 1.77 & 1.48 & 1.24 & 12.18 & 3.40 & 2.01 & 10.47 & 10.72 \\
48 & 46.03 & 5.43 & 126.0 & 4.318 & 0.954 & 0.054 & 0.232 & 0.012 & 0.054 & 0.61 & 1.46 & 1.16 & 1.11 & 13.35 & 3.12 & 1.97 & 10.76 & 10.97 \\
		\hline
	\end{tabular}
\end{table*}
\setlength\tabcolsep{6pt}

We now turn to a detailed, statistical comparison of our simulation results with extant observations.
As noted in \citet{2015aBattaia}, 
much of the available literature on LABs and similar objects report 
different imaging techniques (e.g. slit or 2-dimensional) 
and of different regions (e.g. extended emission or core).
Thus it is useful to bear in mind the heterogeneity in assessing the comparison results.

In the deepest \heii and \civ narrow-band images ever taken,
\citet{2015aBattaia} fail to detect \heii or \civ emission above the 1$\sigma$ surface brightness (SB) 
detection limit of $4.2 \times 10^{-19} 
{\rm \: and \:} 6.8 \times 10^{-19} {\rm \: ergs \: s^{-1} \: cm^{-2} \: arcsec^{-2}}$, respectively,
in two-dimensional images of 13 LABs. 
They are able to place upper limits on the ratios of these two lines to \lya, with the strongest 
constraints of \heii/\lya\ $<0.11$ and \civ/\lya\ $<0.16$ due to the two LABs with the largest isophotal area.
The 2$\sigma$ SB threshold for \lya\ is 
$2.2 \times 10^{-18}$ ergs s$^{-1}$ cm$^{-2}$ arcsec$^{-2}$ \citep{2004Matsuda}. 
\citet{2009Prescott} obtain slit spectra of the emission lines with a 1$\sigma$ detection limit of $\sim10^{-17}$
ergs s$^{-1}$ cm$^{-2}$ arcsec$^{-2}$, and determine line ratios for \heii/\lya\ and \civ/\lya\ from observations of a giant 
\lya\ nebula (PRG1). \citet{2013Prescott} study two other nebulae (PRG2, PRG3) with 1$\sigma$ detection threshold 
of $\sim10^{-18} {\rm \: ergs \: s^{-1} \: cm^{-2} \: arcsec^{-2}}$. 
\citet{2005Dey} also performed observations of a LAB with a SB detection limit of $\sim10^{-18}$
ergs s$^{-1}$ cm$^{-2}$ arcsec$^{-2}$. 
\citet{1991bHeckman} perform analysis of slit spectra of several radio-loud quasars (QSRs),
which have similar properties compared to 
LABs with respect to size and extent of the \lya\ emission. The detection limit of the observations was approximately $3.0 \times 
10^{-17}$ ergs s$^{-1}$ cm$^{-2}$ arcsec$^{-2}$ \citep{1991aHeckman} and they obtain upper limits on the emission line ratios.

In order to place the observational results detected with varying surface brightness detection limits,
we have adjusted all observed data to 
the \lya\ surface brightness threshold $2.2 \times 10^{-18} {\rm \: ergs \: s^{-1} \: cm^{-2}}$ ${\rm \: arcsec^{-2}}$ 
of \citet{2004Matsuda}. 
This adjustment process is calibrated as follows: We use radiative transfer calculations for the galaxies in our sample (listed in 
Table~\ref{tab:ratios}), which allows us to compute ratios of fluxes of the \lya\ line at any two imposed surface brightness thresholds. We then average over the 40 galaxies to obtain the mean of the appropriate ratio in question
between two adopted surface brightness thresholds.
Since our calculations of three lines here do not involve detailed radiative transfer, which are
important for both \lya\ and \civ lines,
we also apply the same adjustments to the simulated galaxies, but on an individual galaxy basis. The adjustment 
consists of multiplying the \heii/\lya\ emission line ratios by a ratio of \llya values at the observed and 
desired surface brightness thresholds. We do not apply this adjustment to \civ/\lya\ since their 
emissions track each other.
We choose our method of calibrating various observations of varying SB 
thresholds to a common one, because it allows us to place all data, observational or 
theoretical, on the same plot, instead of making comparisons between simulation and each 
observational data set on separate plots. Given the relatively small sample size, this is 
preferred also for making statistical tests, such as the KS test we perform.

Figure~\ref{fig:ratios} shows a quantitative comparison between our simulation results and observations.
First, it is evident that the upper limits by \citet{2015aBattaia} are visually consistent with our simulation results
in that the bulk of the simulated results are below the upper limits in both x and y axes.
There is no significant dependence on mass, except perhaps that lower mass galaxies 
represent the smallest ratios, as shown by the predominantly blue diamonds in the lower left corner
of Figure~\ref{fig:ratios}.
Second, a comparison between our simulations results and 
the observed detections (not counting the upper limits), using a two-sample Kolmogorov-Smirnov test, 
yields a p-value of 0.24 for the \heii/\lya\ ratio data and 0.13 
for the \civ/\lya\ ratio data. 
These two KS p-values, if treated as independent,
would give a combined KS p-value ($=p_1p_2[1-\ln(p_1p_2)]$) of 0.14, which indicates an acceptable match.
We find that we could adjust the parameter $\beta$ 
in Eq~\ref{eq:absorption} down somewhat to significantly increase 
the KS test p-value for the \heii/\lya\ ratio;
the KS test p-value for the \civ/\lya\ ratio has not been optimized either.
This is because we feel that such an exercise is not necessarily the most meaningful
and may lead to a misperception of too good agreement between simulations and observations.
Such an exercise will become useful, only when the number of observational detections 
increases by an order of magnitude.
In short, our simulation results with two parameters provide, for the first time,
a successful model to reproduce the two line emission ratio 
\heii/\lya\ and \civ/\lya\ in LABs, in addition to simultaneously 
reproducing the \lya\ luminosity function and luminosity-size relation, as shown earlier in \citet{2013dCen}.

\section{Underlying Physics in Our Model for the Emission Lines}

\ifmnras

\begin{figure}
\centering
	\includegraphics[width=83mm]{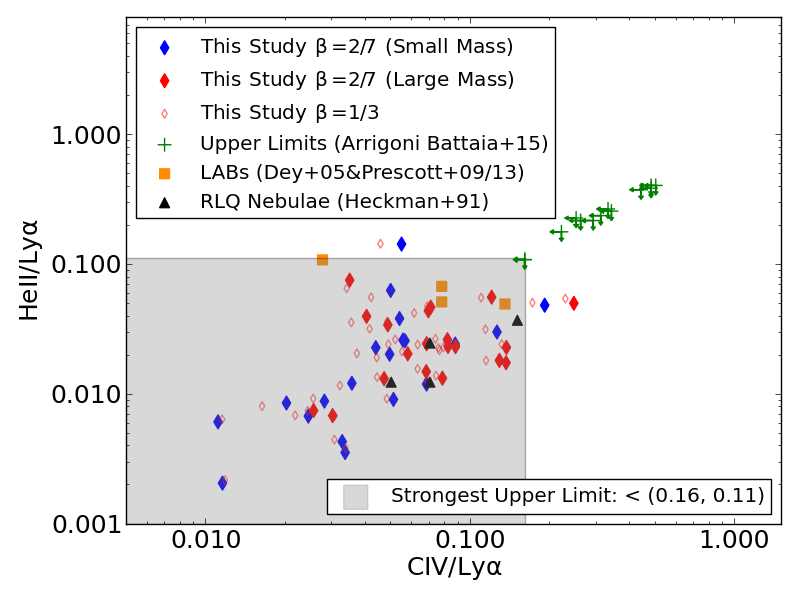}
    \caption{Comparison of the model line ratios (blue and red diamonds) with observations.
Blue and red divide our galaxy sample into two halves, lower 
($5.39 \times 10^{10}$ - $7.76 \times 10^{10}$M$_{\odot}$) and 
upper ($9.17 \times 10^{10}$ - $4.33 \times 10^{11}$M$_{\odot}$) stellar mass bins respectively.
For comparison, we also plot ratios from a varied $\beta$ (red open diamonds).
The observational data are detections from \citet{1991bHeckman} (solid black triangles)
of several radio-loud quasars,
from \citet{2005Dey}, \citet{2009Prescott} and \citet{2013Prescott} (solid orange squares) 
of LABs,
and upper limits from \citet{2015aBattaia} (green pluses with arrows),  
with the strongest upper limits shown as the grayed shaded region.
}
    \label{fig:ratios}
\end{figure}

\else

\begin{figure}[!ht]
\centering
	\includegraphics[width=150mm]{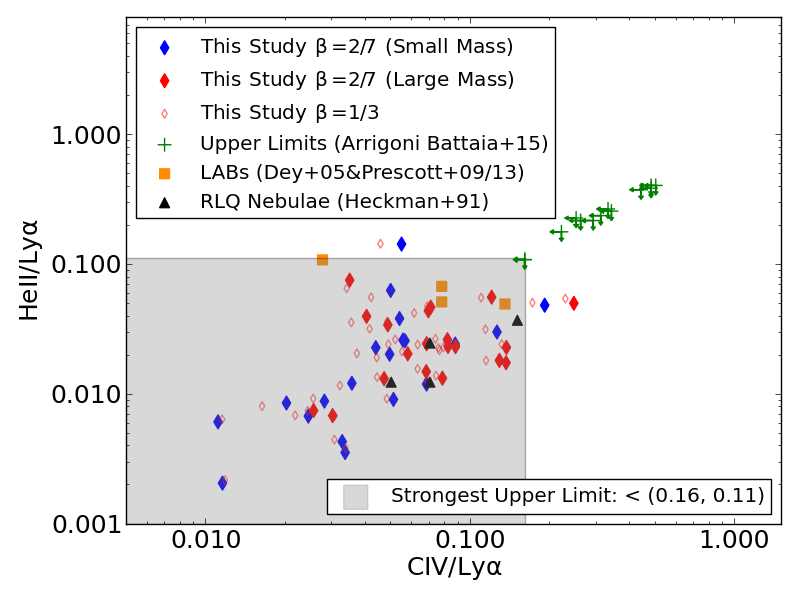}
    \caption{Comparison of the model line ratios (blue and red diamonds) with observations.
Blue and red divide our galaxy sample into two halves, lower 
($5.39 \times 10^{10}$ - $7.76 \times 10^{10}$M$_{\odot}$) and 
upper ($9.17 \times 10^{10}$ - $4.33 \times 10^{11}$M$_{\odot}$) stellar mass bins respectively.
For comparison, we also plot ratios from a varied $\beta$ (red open diamonds).
The observational data are detections from \citet{1991bHeckman} (solid black triangles)
of several radio-loud quasars,
from \citet{2005Dey}, \citet{2009Prescott} and \citet{2013Prescott} (solid orange squares) 
of LABs,
and upper limits from \citet{2015aBattaia} (green pluses with arrows),  
with the strongest upper limits shown as the grayed shaded region.
}
    \label{fig:ratios}
\end{figure}

\fi

The combination of mass budget constraint and required high \lya\ luminosity
place strong physical constraints on volumetric density, column density and size
of the gas in the extended region around the quasar UM287 observed
in the photoionization fluorescence model of \citet{2015bBattaia}: 
${\rm n_{H}} \gtrsim 3  {\rm \: cm^{-3}}, N_{{\rm H}} \lesssim 10^{20}
{\rm \: cm^{-2}}$, and gas clumps with $R \lesssim 20  {\rm \: pc}$. 
They find that, 
unless the metallicity of the line emitting dense gas is very low ($\lesssim 0.001\zsun$),
the expected \heii/\lya\ ratio values obtained in their model reside above the 
strongest upper limits (the left-most green plus in Figure~\ref{fig:ratios}).
Higher metallicities ($0.1 - 1.0\zsun$) are permitted by 
only if $\log {(\rm n_{H}/cm^{-3})} \sim 1.5$.
While not direct proof of the relatively high metallicity for the \heii emitting gas,
we find in the simulation that the emission-weighted metallicity of the \heii emitting gas
is $\sim 1.1\zsun$, ranging from $0.37 - 1.77\zsun$. Even though some galaxies have 
a very high emission-weighted density, the majority of our sample is below
$10 \: {\rm cm}^{-3}$.

Our model is based on {\it ab initio} ultra-high resolution cosmological hydrodynamical simulations
and takes into account all possible emission processes. 
This includes gravitational shocks due to collapse of structures (galaxies within proto-clusters)
and feedback shocks due to supernova explosions, which are both energetically important,
in addition to photoionization by stars.
Although an additional component of radiation from the central AGN may be included,
we do not do so in this analysis and, as we have shown, it is not necessarily needed or beneficial in terms
of providing a good model matching observations.

\ifmnras

\begin{figure*}
\setlength\tabcolsep{3pt}
	\begin{tabular}{ccc} 
		\includegraphics[trim={0.2cm 0 5.8cm 0},clip,width=55.8mm]{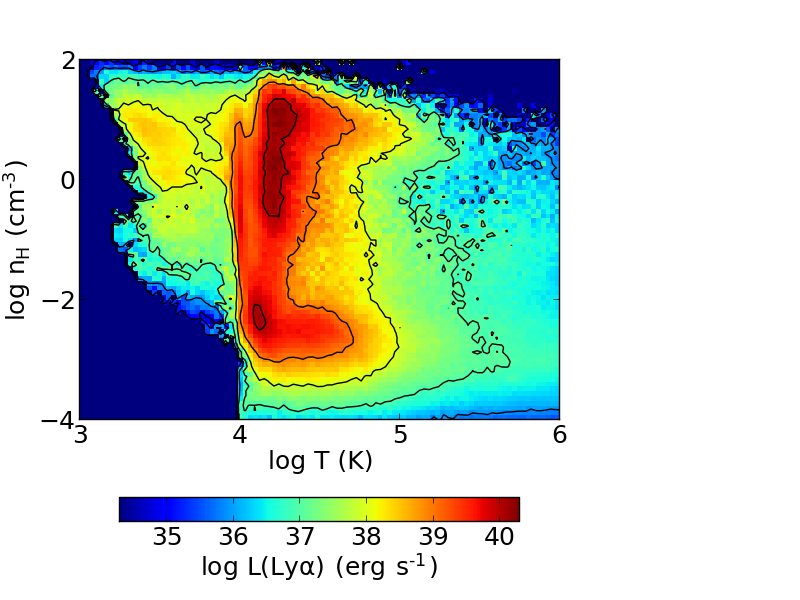}
		&
		\includegraphics[trim={0.2cm 0 5.8cm 0},clip,width=55.8mm]{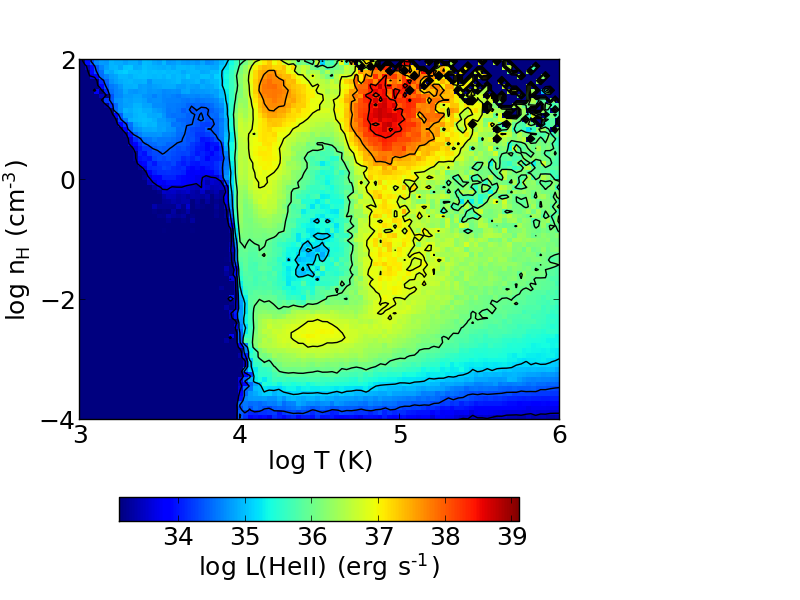}
		&
		\includegraphics[trim={0.2cm 0 5.8cm 0},clip,width=55.8mm]{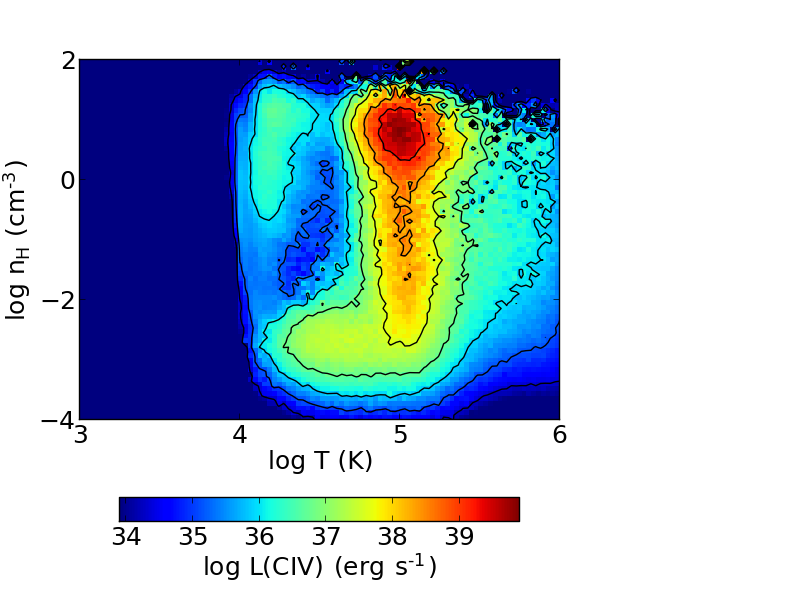}
		\\
    \end{tabular} 
    	\begin{tabular}{cc} 
		\includegraphics[width=90mm]{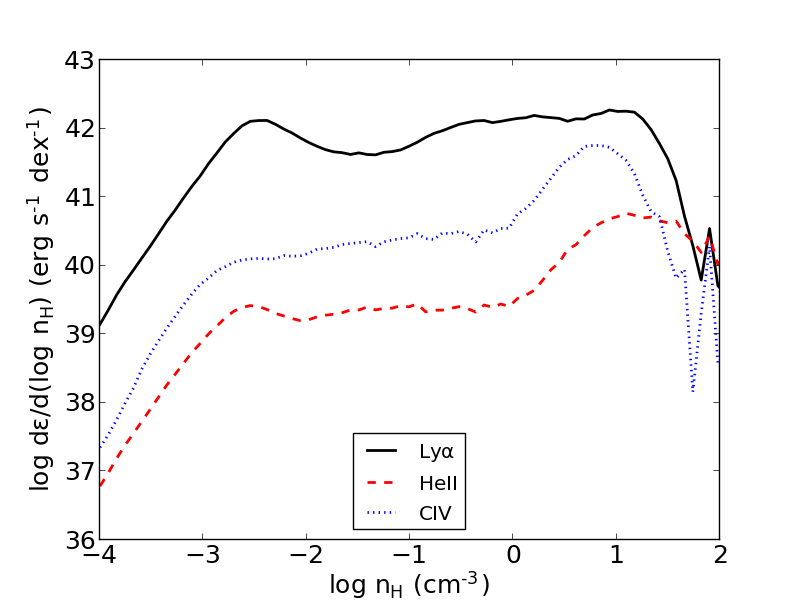}
		&
		\includegraphics[width=90mm]{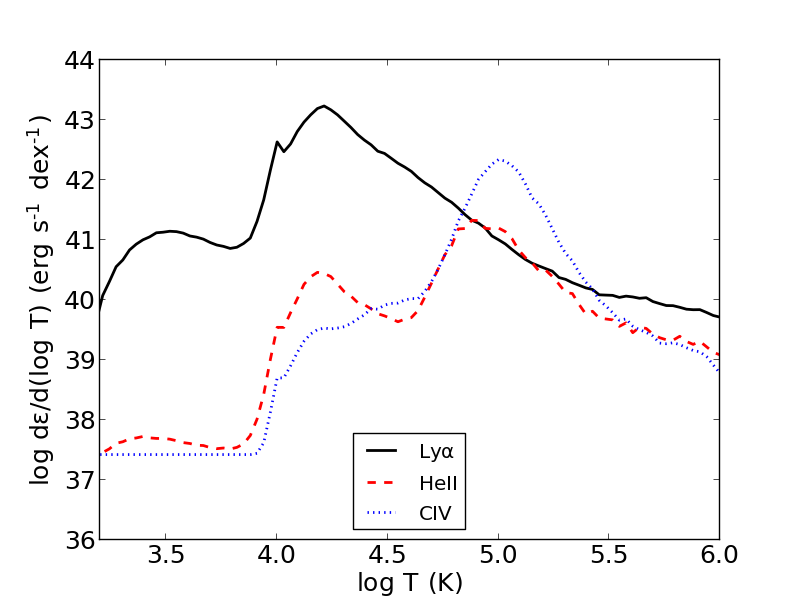}
		\\
    \end{tabular}
    \caption{
Top row: emission of each of the three lines (left panel for \lya, middle panel for \heii and right panel for C\,{\scriptsize IV}) 
in the temperature-density phase diagram.
Bottom row: the distribution of each emission line in density (left panel) and temperature (right panel)
for the \lya, \heii and \civ lines shown in black solid, red dashed and blue dotted curves, respectively,
averaged over the 40 galaxies.
}
    \label{tab:phase}
\setlength\tabcolsep{6pt}
\end{figure*}

\else

\begin{figure}[!ht]
\setlength\tabcolsep{3pt}
	\begin{tabular}{ccc} 
		\includegraphics[trim={0.2cm 0 6cm 0},clip,width=55.8mm]{Hlya1216TemperatureDensity.png}
		&
		\includegraphics[trim={0.2cm 0 6cm 0},clip,width=55.8mm]{HeII1640TemperatureDensity.png}
		&
		\includegraphics[trim={0.2cm 0 6cm 0},clip,width=55.8mm]{CIVtotalTemperatureDensity.png}
		\\
    \end{tabular} 
    	\begin{tabular}{cc} 
		\includegraphics[width=90mm]{total_Density_sideplotDATA.png}
		&
		\includegraphics[width=90mm]{total_Temperature_sideplotDATA.png}
		\\
    \end{tabular}
    \caption{
Top row: emission of each of the three lines (left panel for \lya, middle panel for \heii and right panel for C\,{\scriptsize IV}) 
in the temperature-density phase diagram.
Bottom row: the distribution of each emission line in density (left panel) and temperature (right panel)
for the \lya, \heii and \civ lines shown in black solid, red dashed and blue dotted curves, respectively,
averaged over the 40 galaxies.
}
    \label{tab:phase}
\setlength\tabcolsep{6pt}
\end{figure}

\fi

The physical characteristics of the emission regions in our model are shown in Figure~\ref{tab:phase}. 
The emission of each of the three lines (left panel for \lya, middle panel for \heii and right panel for C\,{\scriptsize IV}) 
in the temperature-density phase diagram is shown in the top row,
whereas the distributions of each emission line in density (left panel) and temperature (right panel)
for \lya, \heii and \civ lines are shown in black solid, red dashed and blue dotted curves, respectively,
in the bottom row.
We see in Figure~\ref{tab:phase} that
most of the \lya\ emission originates from the gas with $\log {(\rm n_{H}/cm^{-3})}=0-1$ 
and $\log {(\rm T/K)} \sim 4.2$. 
Since \lya\ is a resonant line, the most efficient (and economical) powering mechanism 
via collisions is normally collisional excitation, instead of collisional ionization.
Thus, a most noteworthy fact is the temperature of $10^{4.2}$K for the \lya\ emitting gas,
which is the optimal temperature for \lya\ line emission via excitation, where 
the fractions of neutral and ionized hydrogen are roughly comparable.
It is thus clear that collisional excitation (rather than collisional ionization) by combined gravitational and feedback shocks
is a major source of \lya\ emission.
From the cooling point of view, this indicates that \lya\ line emission is a significant cooling mechanism
for shock heated gas.
The metallicity ($\sim 0.5 Z_{\odot}$) of the \lya\ emitting gas 
in the range of $\log {(\rm n_{H}/cm^{-3})}=0-1$ (Table~\ref{tab:ratios})
provides a further differentiating factor
and indicates that feedback shocks are the likely dominant powering source for the collisional excitational generated
\lya\ emission in LABs. 
The small locus at density of $\sim 10^{-3}$cm$^{-2}$ and temperature of $\sim 10^4$K
in the top left panel of Figure~\ref{tab:phase}
is due to photoionization produced \lya\ emission (via recombination), the other primary power source. The photoionization seems to be only on 
halo-scales as seen in Figure~\ref{fig:prof}.
The separation of the collisionally powered and photoionization powered for the \lya\ line
is easily visible in the bottom-left panel of Figure~\ref{tab:phase}.

Both \heii and \civ emission each have a minor photoionization component, seen as 
the loci at hydrogen density of $\sim 10^{-3}$cm$^{-3}$ 
and temperature of $\sim 10^{4-5}$K in the top middle and bottom right panels.  
In stark contrast to the \lya\ line, the photoionization powered emission for \heii\ and \civ\ lines
is much weaker, seen in the lack of significant emission peaks at the low density end
($\sim 10^{-3}-10^{-2}$cm$^{-3}$) in the red dashed and  blue dotted curves 
in the bottom-left panel of Figure~\ref{tab:phase}.
Note that the temperature of $\log {\rm T}\sim 5$ is the temperature for maximum
abundance of \civ by collisional processes and thus is consistent with collision excitation being the responsible process
for \civ emission.
Indeed, this conclusion is reached by \citet{2015aBattaia} and \citet{2015bBattaia}; the high sensitivity 
of this line to temperature is noted in \citet{2004Groves}.
This is also consistent with the fact that the concerned \civ emission line is a resonant line as well.
Most of the \heii emission originates from regions of similar temperature ($\log {\rm T}\sim 5$)
as the \civ emission, but somewhat more dense (${\rm n_H}=5-20$cm$^{-3}$).
It is noted that at $\log {\rm T}=5$ the dominant ionization state of helium is He III.
Thus, the dominant emission process for \heii line is recombination via collisional ionization of He II.
This is consistent with the \heii line being a non-resonant line 
(note that the \heii 1640\AA\ line is in analogy to hydrogen H$\alpha$ line).

To better understand the variations and commonalities between galaxies 
with respect to the emission of the three lines concerned,
we list in Table~\ref{tab:ratios} the emission-weighted metallicity, density and temperature for the three lines
for the 40 simulated galaxies analyzed.
We see that the metallicity ranges from $0.28$ to $1.77\zsun$.
There is no universal trend of the metallicity with respect to the three lines,
although in a typical case the mean metallicity increase from \lya\ to \civ significantly and then to \heii\ slightly.
The typical range of (\lya, \civ, \heii) emission weighted metallicities
are ($0.5\pm 0.2$, $1.0\pm 0.2$, $1.1\pm 0.3)\zsun$.
This is likely due to metal enrichment and mixing in the CGM \citep{2010Shen,2013Crain,2014Brook,2016Ford},
which is accounted for in our simulation by
following metal transport in a spatially resolved
fashion at the high resolution of the AMR grid.

It is found that in most cases
the \lya\ emission-weighted gas density is in the range of $0.5-3$cm$^{-3}$, 
comparable to but slightly lower than $2-5$cm$^{-3}$ for \civ, 
which in turn is slightly lower than $5-20$cm$^{-3}$ for \heii.
It should be noted that this general trend does not hold on an individual basis,
neither is there a definitive correlation between stellar mass of the galaxy and the emission-weighted gas density.
This is indicative of and consistent with the notion that the gas density, 
the gas that bears the physical characteristics for strong line emission for three lines concerned in particular,
fluctuates in time.
This bodes well with the notion that star formation, which is fueled by gas accretion,
may fluctuate strongly and hence bursty in nature.
Finally, the emission-weighted temperatures of the three lines
differ significantly with that of \lya\ slightly above $10^4$K,
that of both \civ\ and \heii\ about $10^5$K. 
The emission-weighted temperatures of the \civ\ line 
is very close to that of the \heii\ line,
visible in the red dashed and  blue dotted curves 
in the bottom-right panel of Figure~\ref{tab:phase},
although the \heii\ line has a more pronounced photoionization peak at $T\sim 10^{4.2}$K.
Taking into account the differences in metallicity, temperature and density
of gas emitting the three emission lines,
it is clear that, while there may be some overlaps, especially between \heii\ and \civ\ lines,
the zero-order picture emerging is that
the three lines largely originate from different gas regions in a multi-phase interstellar and
circumgalactic medium that is shock-heated primarily by stellar feedback/supernovae
and secondarily by gravitational energy of accreting gas.
Figure~\ref{tab:slices} demonstrates 
this finding through emission weighted maps of one of the galaxies in our sample. 
The emission regions trace different gas properties 
(temperature, density and metallicity) with some degree of overlap.

While our model appears to have the ability to account for the observed line ratios
of \heii/\lya\ and \civ/\lya, based on a model that has provided the first 
successful explanation for \lya\ emission properties 
(size-luminosity relation and luminosity function) of observed LABs,
it is useful to bear in mind that there is still significant freedom available to us 
for adjustments of the two parameters, $\beta$ and $\gamma$ in Eq~\ref{eq:absorption},
largely due to our inability to precisely constrain simultaneously the 
distributions of gas density, temperature and metallicity, upon which dust formation and hence the obscuration/absorption
sensitively depends, among others.
This ambiguity may be somewhat mitigated by the fact that
the number of parameters in our modeling is outnumbered by the number of observational constraints.
Nevertheless, we plan to investigate other diagnostic emission and absorption lines 
to further test the consistency of our model.

\section{Discussion and Conclusions}

\ifmnras

\begin{figure}
\setlength\tabcolsep{3pt}
	\includegraphics[trim={0 0 0 0},clip,width=83.0mm]{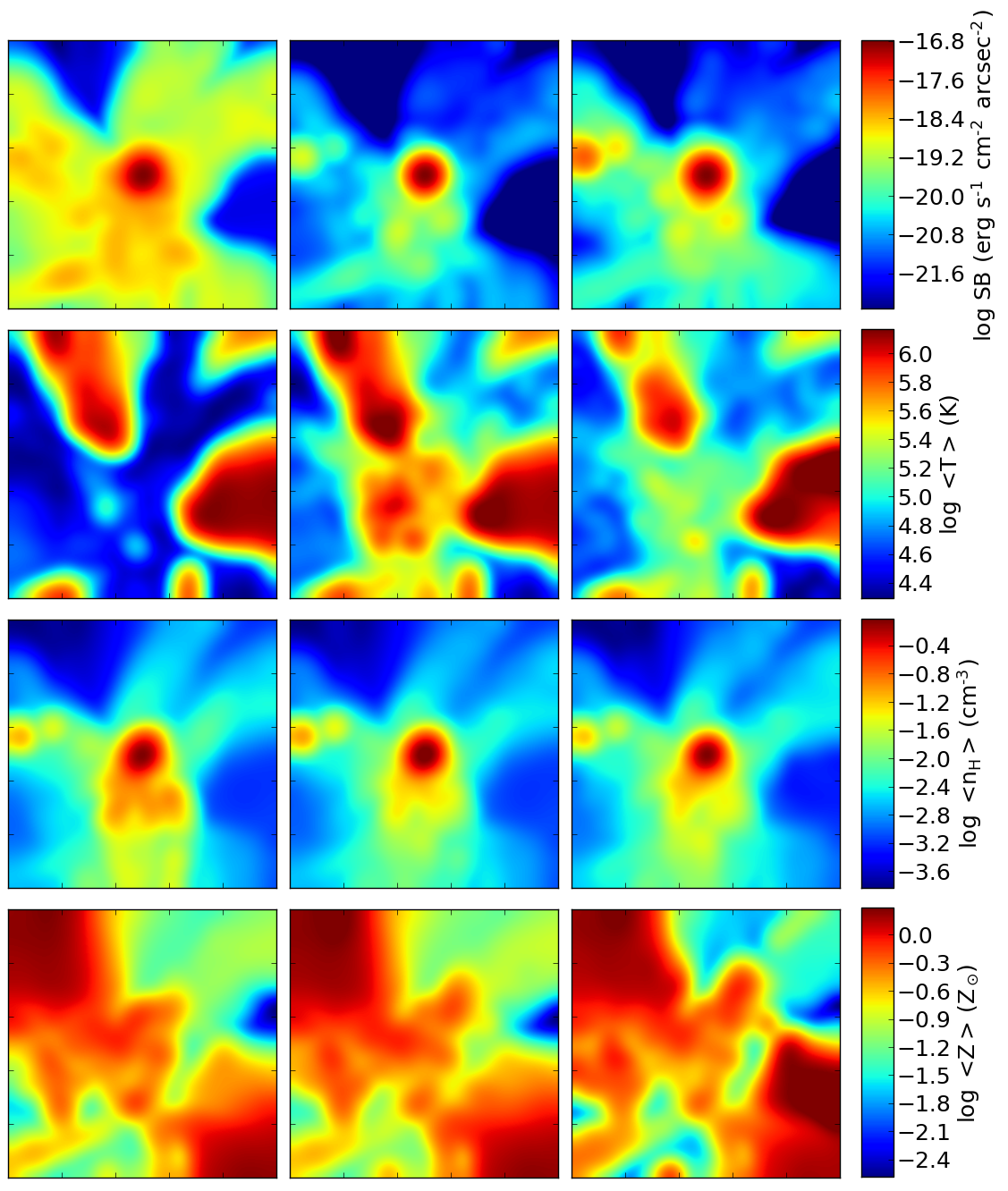}
    \caption{
Projection plots (100 kpc x 100 kpc) for a randomly selected galaxy (ID 5). 
From left to right, emission lines  
\lya, \heii, and C\,{\scriptsize IV}. From top to bottom, surface brightness, and emission weighted temperature, density, and metallicity.
Distinct emission regions for each of the three lines are visible, and are characteristic of different combinations 
of properties.
}
    \label{tab:slices}
\setlength\tabcolsep{6pt}
\end{figure}

\else

\begin{figure}
\setlength\tabcolsep{3pt}
	\includegraphics[trim={0 0 0 0},clip,width=150.0mm]{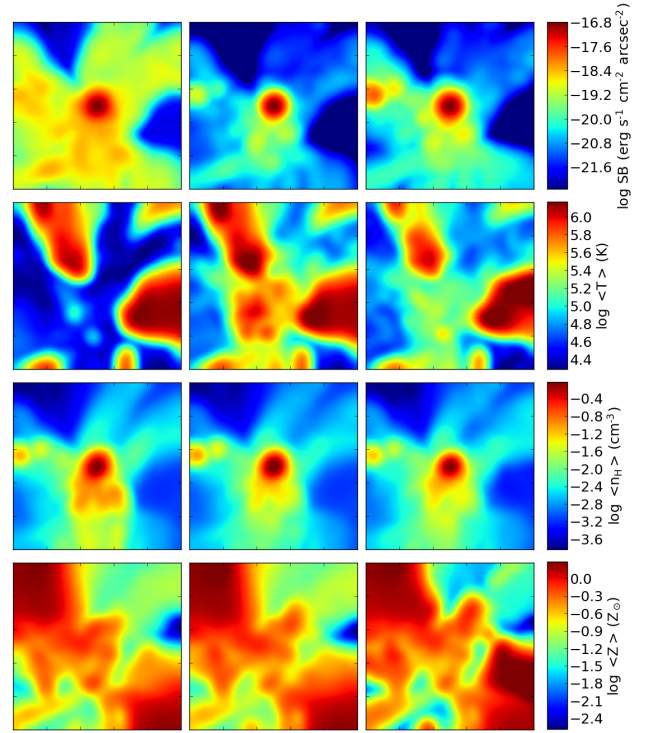}
    \caption{
Projection plots (100 kpc x 100 kpc) for a randomly selected galaxy (ID 5). 
From left to right, emission lines  
\lya, \heii, and C\,{\scriptsize IV}. From top to bottom, surface brightness, and emission weighted temperature, density, and metallicity.
Distinct emission regions for each of the three lines are visible, and are characteristic of different combinations 
of properties.
}
    \label{tab:slices}
\setlength\tabcolsep{6pt}
\end{figure}

\fi

Utilizing {\it ab initio} ultra-high resolution ({\color{red}\bf LAOZI}) hydrodynamical simulations, 
we have previously shown that the observed size-luminosity and luminosity function of 
LABs can be successfully reproduced, with the extended Ly$\alpha$ 1216\AA\ 
emission powered by a combination of central star formation 
and central shock heated gas \citep{2013dCen}.
In this study, we investigate two additional emission lines of LABs: 
\heii 1640\AA\ and \civ 1549\AA. 
Two parameters are introduced and adjusted for modeling the dust attenuation.   
With that, we show that our simulations can reproduce simultaneously 
the observed \civ/\lya\ and \heii/\lya\ ratios,
in addition to the agreements for the LAB luminosity function, luminosity-size relation that have been achieved before.

We show that the \heii 1640\AA\ and \civ 1549\AA\ are largely powered by shocked heated gas,
due primarily to feedback shocks from supernovae and secondarily to gravitational gas accretion shocks.
The \civ\ emission line is primarily
collisional excitation driven, while
the \heii\ emission line is powered mainly by collisional ionization,
with in situ photoexcitation/photoionization being a minor contributor for either of the lines.
The \lya\ emission line is powered by a combination of stellar radiation and collisional excitation, both being 
significantly more centrally concentrated than the observed \lya\ emission surface brightness profile;
we have shown in \citet{2013dCen} that resonant scattering and spatial diffusion of \lya\ photons
produce the extended, diffuse emission observed, in conjunction with clustering of galaxies.
\citet{2013dCen} also show
that the peak of the \lya\ emission need not originate from a compact source via detailed radiative transfer 
methods, consistent with the findings of \citet{2012Prescott} and \citet{2015Prescott}.

We find that the density, temperature and metallicity of the gas responsible for each emission line
is significantly distinct, in a multi-phase interstellar and circumgalactic medium.
We see in Figure~\ref{tab:phase} that
most of the \lya\ emission originates from the collisionally excited gas with $\log{(\rm n_{H}/cm^{-3})}=0-1$, 
temperature $\log {(\rm T/K)} \sim 4.2$ and subsolar metallicity.
Most of the \heii emission originates from regions of temperature of $\log {\rm T}\sim 5$,
density of ${\rm n_H}=5-20$cm$^{-3}$ and solar metallicity. 
Overall, the physical properties of the \civ emitting gas is similar to those of the \heii\ emitting gas.
On an individual galaxy basis, however, the mean emission-weighted 
gas densities, temperatures and metallicities for \civ\ and \heii\ lines
differ substantially, indicative of 
significantly differing, albeit overlapping, emission regions.

We have used the very useful and versatile
analysis software yt version 2.6 \citep{2011Turk}
for some of our analysis.
Computing resources were in part provided by the NASA High-
End Computing (HEC) Program through the NASA Advanced
Supercomputing (NAS) Division at Ames Research Center.
The research is supported in part by NASA grant NNX11AI23G,
NASA grant NNX14AC89G and NSF grant AST-1208891.




\bibliographystyle{mnras}
\bibliography{astro} 






\label{lastpage}
\end{document}